\documentclass[reprint,amssym,amsmath,prb]{revtex4-1}
\pdfoutput=1
\usepackage{amsmath,amssymb,amsfonts,upgreek}
\usepackage{graphicx}

\usepackage{booktabs}

\newcommand{\btfo}{\ensuremath{\mathrm{Bi_5FeTi_3O_{15}}}}
\newcommand{\bio}{\ensuremath{\mathrm{Bi_2O_2}}}
\newcommand{\bi}{\ensuremath{\mathrm{Bi^{3+}}}}
\newcommand{\fe}{\ensuremath{\mathrm{Fe^{3+}}}}
\newcommand{\ti}{\ensuremath{\mathrm{Ti^{4+}}}}
\newcommand{\lowsym}{\ensuremath{A2_1am}}

\begin{document}

\title{Controlling the cation distribution and electric polarization
  with epitaxial strain in Aurivillius-phase Bi$_5$FeTi$_3$O$_{15}$}

\author{Axiel Ya\"el Birenbaum} 
\email{yael.birenbaum@mat.ethz.ch} 
\affiliation{Materials Theory, ETH Z\"urich, Wolfgang-Pauli-Strasse
  27, 8093 Z\"urich, Switzerland} 
\author{Claude Ederer} 
\email{claude.ederer@mat.ethz.ch} 
\affiliation{Materials Theory, ETH Z\"urich, Wolfgang-Pauli-Strasse
  27, 8093 Z\"urich, Switzerland}

\date{\today}

\begin{abstract}
This work explores the impact of in-plane bi-axial (epitaxial) strain
on the cation distribution and electric polarization of the
Aurivillius-phase compound \btfo\ using first-principles electronic
structure calculations. Our calculations indicate that the site
preference of the \fe\ cation can be controlled via epitaxial strain.
Tensile strain enhances the preference for the \emph{inner} sites
within the perovskite-like layers of the Aurivillius-phase structure,
whereas compressive strain favors occupation of the \emph{outer}
sites, i.e., the sites close to the \bio\ layer. Controlling the
distribution of the magnetic cations offers the possibility to control
magnetic order in this magnetically dilute system. Furthermore, the
magnitude of the electric polarization is strongly strain-dependent,
increasing under tensile strain and decreasing under compressive
strain. We find strongly anomalous Born effective charges, both of the
\bi\ and the \ti\ cations.
\end{abstract}

\maketitle


Controlling the properties of complex transition metal oxides by
epitaxial strain, i.e., by growing thin films of a certain material on
a substrate with specific lattice mismatch, has emerged as a very
efficient way for designing optimized
functionalities.~\cite{Schlom_et_al:2008} In particular, the effect of
epitaxial strain on the ferroelectric properties of perovskite
materials is well studied,~\cite{Schlom_et_al:2007} and dramatic
enhancements of polarization and ferroelectric ordering
temperatures,~\cite{Choi_et_al:2004} as well as emergence of
ferroelectricity in otherwise nonpolar materials have been
reported.~\cite{Haeni_et_al:2004,Bhattacharjee/Bousquet/Ghosez:2009}

Recently, layered perovskite-related systems have come into focus as
being potentially more amenable to developing polar lattice
distortions compared to bulk perovskites.~\cite{Benedek_et_al:2015}
Examples include artificial perovskite superlattices and double
perovskites, as well as several families of naturally-layered
perovskite-derived crystal structures such as the Ruddlesden-Popper
series, Aurivillius-phases, or Dion-Jacobson systems. Only few studies
addressing the strain response of these naturally-layered materials
are currently available. Such studies are, however, of great interest
due to the different mechanism underlying the ferroelectricity in
these systems, which could lead to a different strain response
compared to bulk perovskite ferroelectrics.

\begin{figure}
\centering
\includegraphics[width=0.7\columnwidth]{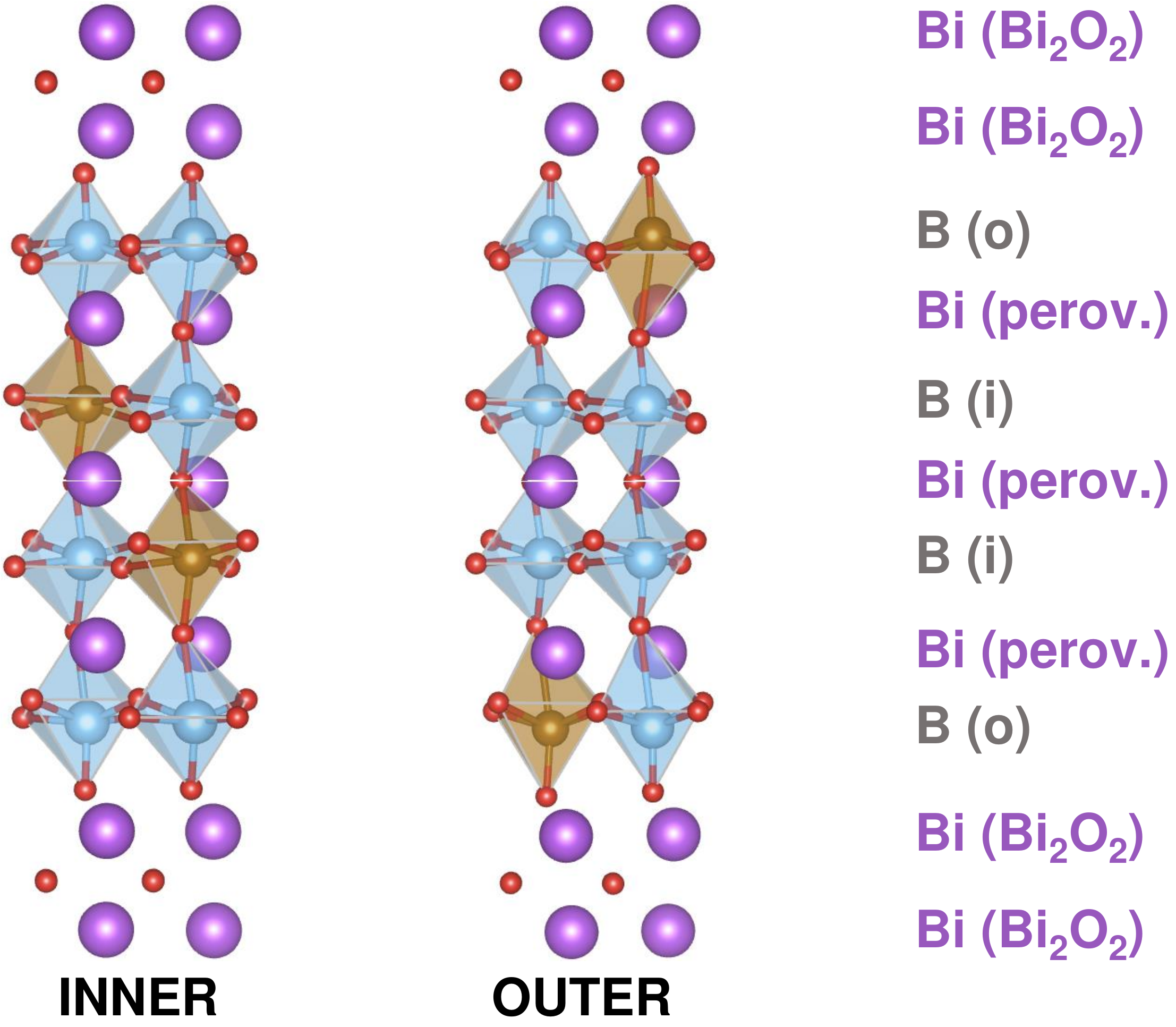}
\caption{(Color online) The two $B$-site cation distributions used in
  this work, \textit{inner} and \textit{outer}. Sites occupied with Fe
  (Ti) are indicated as brown (blue) spheres with corresponding
  coordination octahedra; Bi (O) ions are shown as purple (red)
  spheres. Labels for the cations in the different layers (used for
  the discussion of Born effective charges) are listed on the
  right. This figure was constructed using
  VESTA.\cite{Momma/Izumi:2011}}
\label{fig:config}
\end{figure}

Here we study the case of
\btfo\ (BFTO),\cite{Kubel/Schmid:1992,Hervoches_et_al:2002} a
representative of the family of naturally-layered Aurivillius-phases,
which is of particular interest due to its potential multiferroic
properties.~\cite{Birenbaum:2014bl} The crystal structure of the
Aurivillius-phases consists of $m$ perovskite layers $\mathrm{\left(
  A_{m-1}B_mO_{3m+1} \right)^{2-}}$ stacked periodically along the
[001] direction, and separated by fluorite-like $\mathrm{\left(
  Bi_2O_2 \right)^{2+}}$ layers (see
Fig.~\ref{fig:config}).\cite{Newnham/Wolfe/Dorrian:1971,Frit/Mercurio:1992}
BFTO corresponds to the case with $m=4$.

A previous first-principles study of polarization-strain coupling in
the $m=3$ Aurivillius compound $\mathrm{Bi_4Ti_3O_{12}}$ has shown a
large response of in-plane polarization under bi-axial
strain.\cite{Shah:2010kg} Apart from the introduction of an additional
perovskite layer and the presence of the nominally non-ferroelectric
\fe\ cation, an important additional degree of freedom in BFTO
compared to $\mathrm{Bi_4Ti_3O_{12}}$ is the distribution of \fe\ and
\ti\ cations over the available $B$ sites within the perovskite layers
of the Aurivillius structure. While the \fe/\ti\ cations do not tend
to form an ordered arrangement, a preferential occupation of the inner
perovskite site with Fe has been reported
experimentally~\cite{Lomanova_et_al:2012} and confirmed recently by
density functional theory (DFT) calculations.\cite{Birenbaum:2014bl}
The cation distribution can in principle influence properties such as
electric polarization or the likelihood and character of magnetic
order, and it could also affect the strain response of the material.
Furthermore, a systematic variation of the relaxed lattice constants
with the distribution of Fe over the inner/outer perovskite layers has
been found in Ref.~\onlinecite{Birenbaum:2014bl}, suggesting that it
might be possible to influence the cation distribution by epitaxial
strain. Here, we verify this hypothesis and study the effect of strain
on the cation distribution and on the resulting electric polarization
of BFTO.


We use a unit cell that corresponds to the primitive unit cell of the
experimentally-observed \lowsym\ structure of BFTO, containing two
formula units. There are 10 symmetry-inequivalent ways to distribute 2
Fe and 6 Ti atoms over the 8 available peroskite $B$ sites within this
unit cell.\cite{Birenbaum:2014bl} We focus on two representative
configurations depicted in Fig.~\ref{fig:config}, one with all Fe
sitting in the \emph{outer} perovskite layers and one with all Fe
sitting in the \emph{inner} perovskite layers (denoted as O1 and I1 in
Ref.~\onlinecite{Birenbaum:2014bl}). For both configurations, the
specific distribution of \fe\ and \ti\ cations lowers the
experimentally observed \lowsym\ space group symmetry to monoclinic
$P2_1$.
Both configurations allow to define a centrosymmetric reference
structure with $P2_1/m$ symmetry for calculating the spontaneous
electric polarization by removing all structural distortions that
break inversion symmetry.

In order to model the elastic boundary conditions corresponding to the
epitaxial constraint imposed by a substrate, we fix the two short
in-plane lattice parameters, $a$ and $b$, to be equal to an effective
substrate lattice constant $a'$, and the corresponding lattice vectors
to form a 90$^\circ$ angle. This corresponds to a thin-film/substrate
interface forming a square lattice. Furthermore, for simplicity we
constrain the out-of-plane lattice vector to correspond to a perfect
base-centered orthorhombic Bravais lattice, in spite of the lower
monoclinic symmetry of the whole structure. We find the optimal value
of the out-of-plane lattice parameter $c$ (in our notation
corresponding to the \emph{conventional} orthorhombic Bravais lattice)
by relaxing all ionic positions for fixed $a'$ and different values of
$c$. In the following, whenever comparison with experimental
structural data is made, Ref.~\onlinecite{Hervoches_et_al:2002} is
used, and we define 0\% strain relative to the average experimental
in-plane lattice constant $a_{0}^{\prime} = (a^{\mathrm{exp}} +
b^{\mathrm{exp}})/2 = 5.45 \text{\AA}$.

We perform DFT calculations using the projector augmented wave (PAW)
method as implemented in the Vienna \textit{ab initio} simulation
package
(VASP),~\cite{Kresse/Furthmueller_CMS:1996,Kresse/Joubert:1999} and
the generalized gradient approximation according to Perdew, Burke, and
Ernzerhof optimized for solids (PBEsol).~\cite{Perdew_et_al:2008} Our
PAW potentials include 15 valence electrons for Bi ($6s^2 5d^{10}
6p^3$), 14 for Fe ($3p^6 4s^2 3d^6$), 10 for Ti ($3p^6 4s^2 3d^2$),
and 6 for O ($2s^2 2p^4$). We include a Hubbard ``$+U$'' correction
with $U_\text{eff}=3.0$\,eV to correctly treat the strong interactions
between the Fe $d$ electrons.\cite{Dudarev_et_al:1998} Ionic positions
are relaxed until the residual forces are smaller than $10^{-3}$
eV/\AA. Calculations are converged using a $\Gamma$-centered $k$-point
mesh with $4 \times 4 \times 2$ divisions along the three reciprocal
lattice vectors and a plane wave cutoff energy of 550\,eV.  As shown
previously,\cite{Birenbaum:2014bl} all magnetic couplings in BFTO are
antiferromagnetic, hence we fix antiparallel orientation of the
magnetic moments of the two \fe\ cations within the unit cell.



\begin{figure}
\centering
\includegraphics[width=0.8\columnwidth]{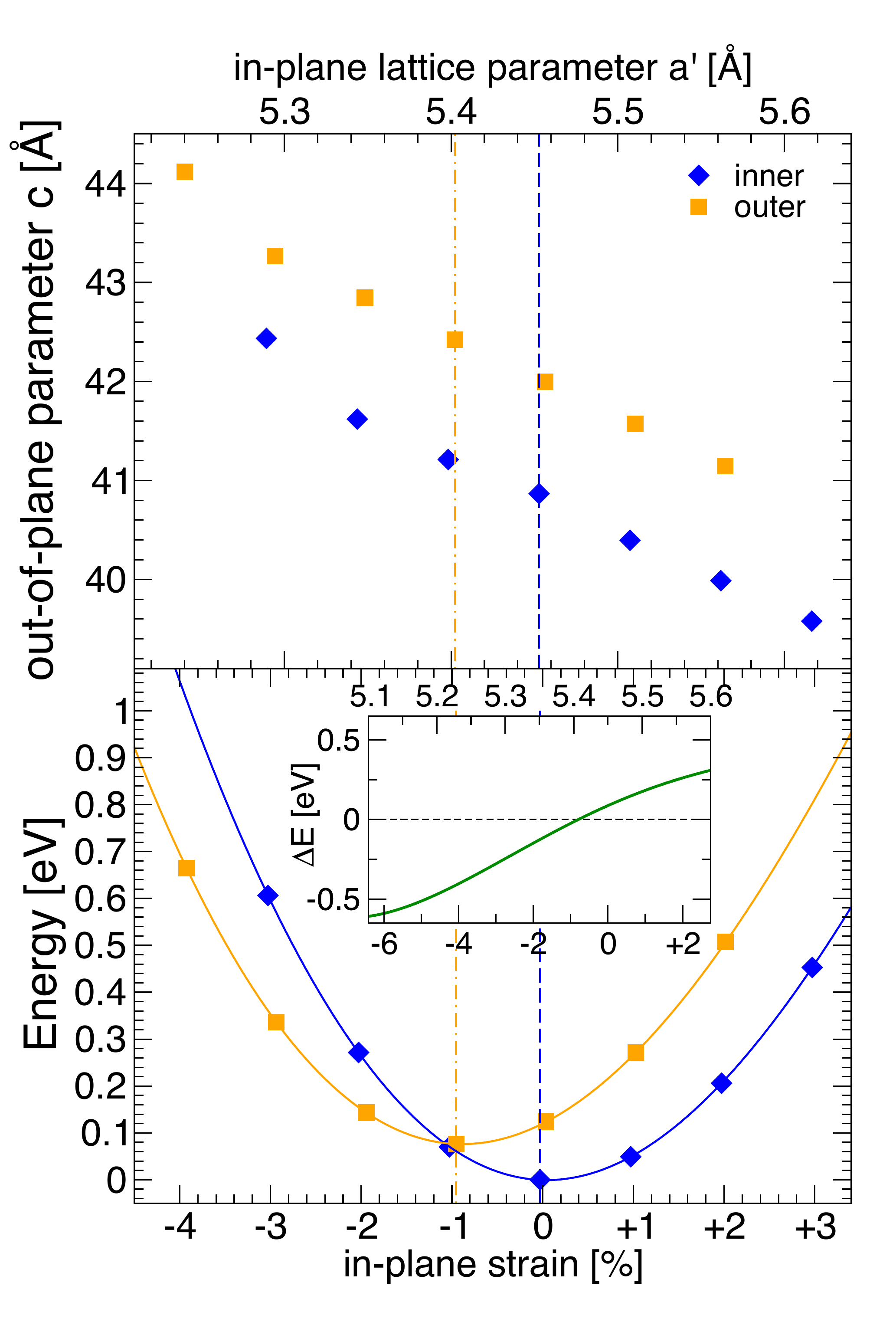}
\caption{(Color online) Out-of-plane lattice parameter $c$ (top) and
  total energy per unit cell (bottom) as a function of in-plane
  biaxial strain (with the axis at the top of the graph giving the
  corresponding in-plane lattice parameter), for the inner (blue
  diamonds) and outer (orange squares) configuration. The dashed lines
  mark the corresponding equilibrium in-plane parameters for the inner
  (blue) and outer (orange) configurations, respectively. The inset
  shows the energy difference $\Delta E$ (per unit cell) between the
  outer and inner configuration for different values of in-plane
  strain.}
\label{fig:strained-lat-pref}
\end{figure}

Fig.~\ref{fig:strained-lat-pref} shows the optimized out-of-plane
lattice parameter $c$ as well as the energy of both inner and outer
configurations for a range of in-plane lattice parameters $a'$.
The minimum of the energy for each configuration indicates the
preferred in-plane lattice parameter for that configuration. Thus, it
can be seen that the inner configuration prefers a larger in-plane
lattice constant $a'$ than the outer configuration and a shorter
out-of-plane lattice parameter $c$, in agreement with the bulk
relaxations presented in Ref.~\onlinecite{Birenbaum:2014bl}. We note
that the position of the energy minimum for the inner configuration,
i.e., the corresponding preferred in-plane lattice parameter, agrees
very well with the averaged experimental in-plane lattice constant,
which serves as our zero strain reference.
For a given in-plane lattice parameter, the outer configuration always
leads to a larger out-of-plane lattice parameter than the inner
configuration. This is due to the strong local tetragonality around
the \fe\ cation occupying an outer site.\cite{Birenbaum:2014bl}
Furthermore, it can be seen that the energy minimum for the inner
configuration is lower than the energy minimum for the outer
configuration, consistent with the inner site preference of \fe\ found
in Ref.~\onlinecite{Birenbaum:2014bl}.

The inner site preference is increased if $a'$ is increased (see
Fig.~\ref{fig:strained-lat-pref}, inset). Thus, tensile strain is
expected to strengthen the inner site preference. On the other hand,
if compressive strain is applied, i.e., for decreasing $a'$, the
energy difference between inner and outer configuration decreases, and
for values below $a' = 5.40$ \AA, corresponding to about $-1$\,\%
strain, the outer configuration has a lower energy than the inner
configuration.

These results imply that the site preference of the \fe\ cation can
indeed be tuned through epitaxial strain. Tensile strain enforces the
preference for the inner sites, while compressive strain can reverse
this preference such that the majority of \fe\ cations occupy outer
sites. We note that these results are obtained by considering only
two, albeit representative, configurations, and that a more complete
treatment would require to consider more configurations. However, the
obtained trend is consistent with the observation from
Ref.~\onlinecite{Birenbaum:2014bl} (and
Fig.~\ref{fig:strained-lat-pref}) that outer configurations generally
favor a more elongated unit cell (along $c$) compared to inner ones.


Next, we address the effect of strain on the electric polarization. We
calculate the spontaneous polarization $P_{\alpha}$ along a given
direction $\alpha$ using Born Effective Charges
(BECs),\cite{Ghosez/Michenaud/Gonze:1998} $Z^{\ast}_{\kappa,\alpha
  \beta}$, as follows
\begin{equation}
P_{\alpha} = \frac{e}{\Omega} \sum_{\kappa,\beta}
Z^{\ast}_{\kappa,\alpha \beta} u_{\kappa,\beta}
\end{equation}
where $\kappa$ indicates the different ions and $u_{\kappa,\beta}$ is
the displacement of ion $\kappa$ along direction $\beta$ between the
paraelectric reference and the relaxed ferroelectric structure. The
BEC tensors are calculated for the polar structures obtained at each
strain value using density functional perturbation theory as
implemented in VASP.

\begin{figure}
\centering
\includegraphics[width=0.85\columnwidth]{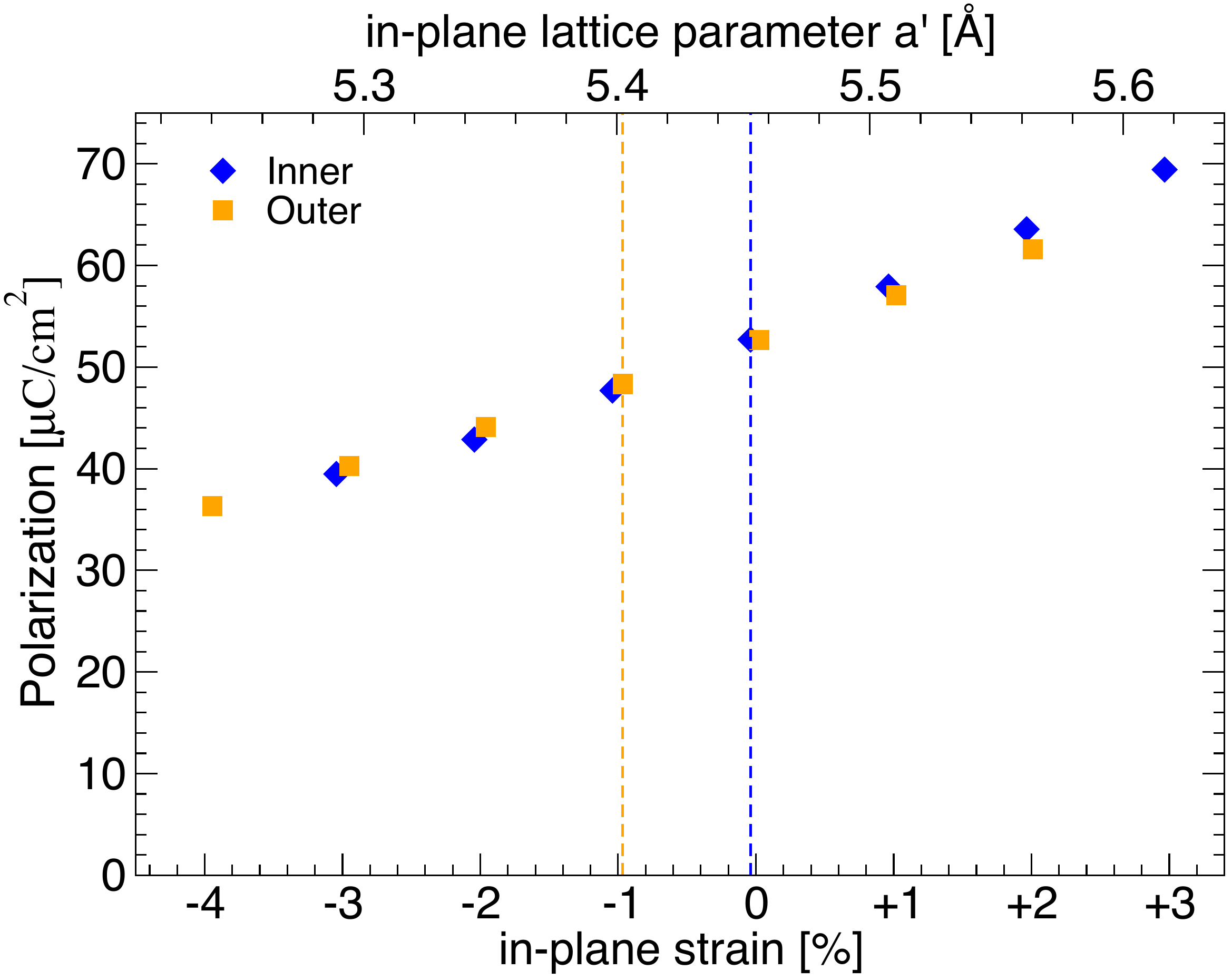}
\caption{(Color online) Spontaneous polarization as function of
  in-plane strain for inner (blue diamonds) and outer (orange squares)
  configurations. The dashed vertical lines indicate the optimal
  in-plane lattice parameters for each configuration.}
\label{fig:strained-pol}
\end{figure}

The calculated polarization as function of strain is depicted in
Fig.~\ref{fig:strained-pol}. By symmetry, the polarization is
restricted to be aligned along the in-plane two-fold screw axis ($b$
axis in standard $P2$ setting, equivalent to the $a$ axis in the
experimental $A2_1am$ structure). At the optimal in-plane lattice
constants of each configuration, we find a spontaneous polarization of
$P=48.3$\,$\mu$C/cm$^2$ for the outer configuration, and
$P=52.7$\,$\mu$C/cm$^2$ for the inner configuration. These results are
consistent with the corresponding bulk values from
Ref.~\onlinecite{Birenbaum:2014bl} ($P=51.5$/$57.9$\,$\mu$C/cm$^2$ for
the outer/inner configuration), calculated using the Berry phase
method.\cite{King-Smith/Vanderbilt:1993} Note that we are using the
BECs calculated for the polar structures to evaluate the spontaneous
polarization. Since the BECs decrease when going from the paraelectric
to the ferromagnetic structure (see Table~\ref{table:bec}), this leads
to a small underestimation of the spontaneous polarization compared to
the exact calculation of $P$ using the Berry phase.

We observe that in-plane tensile strain leads to a strong increase in
polarization and, conversely, compressive strain leads to a strong
decrease. The polarization varies by about $\pm 30$\,\% over the whole
strain region considered here, i.e., from $-$4\% to $+$3\%
strain. This trend is similar to many perovskite ferroelectrics, where
an elongation (compression) along the polar axis increases (decreases)
the ferroelectric displacements of the ions, and thus the
polarization, along that
direction.\cite{Bungaro/Rabe:2004,Dieguez_et_al:2004,Ederer/Spaldin_PRL:2005}
Similar behavior has also been found in previous DFT studies of the
Aurivillius phases Bi$_4$Ti$_3$O$_{12}$ ($m=3$)~\cite{Shah:2010kg} and
SrBi$_2$Ta$_2$O$_9$ ($m=2$).\cite{Yang_et_al:2013}

Furthermore, we can see that for a given in-plane lattice parameter,
the calculated polarization is very similar for both configurations,
in spite of the differences in the out-of-plane lattice parameter $c$
(see Fig.~\ref{fig:strained-lat-pref}). This indicates that the impact
of the Fe/Ti distribution on the electric polarization is minimal. The
difference in spontaneous polarization obtained for the relaxed inner
and outer structures stems mostly from their different in-plane
lattice constants ($a'= 5.45$\,\AA\ and $a'=5.40$\,\AA, respectively).
We can conclude that neither does the presence of the magnetic
\fe\ cations have a negative impact on the overall magnitude of the
electric polarization (compared to, e.g., Bi$_4$Ti$_3$O$_{12}$), nor
does the specific distribution of magnetic cations have a significant
effect on this magnitude.


\begin{table}
\caption{\label{table:bec} BECs (in units of $|e|$) of the different
  cations for the two cation distributions in the corresponding
  centrosymmetric ($P2_1/m$) and polar ($P2_1$) structures. The
  cations are labeled according to Fig.~\ref{fig:config} and separate
  averages are calculated for the in-plane ($xy$) and out-of-plane
  ($z$) diagonal elements of the BEC tensors.}
\begin{ruledtabular}
\begin{tabular}{l|cccc|cccc}
 & \multicolumn{4}{c|}{inner} & \multicolumn{4}{c}{outer} \\
 &\multicolumn{2}{c}{$P2_1/m$} & \multicolumn{2}{c|}{$P2_1$} &\multicolumn{2}{c}{$P2_1/m$} & \multicolumn{2}{c}{$P2_1$} \\ \cmidrule(lr){2-3} \cmidrule(l){4-5}
                & $xy$       & $z$         & $xy$        & $z$    & $xy$       & $z$         & $xy$      & $z$  \\
\hline
\bi(Bi$_2$O$_2$) &  4.95  &  4.81  &  4.81  &  4.85  &  5.08  &  4.51  &  4.84  &  4.42  \\
\bi(perov.)     &  5.66  &  4.08  &  5.00  &  4.50  &  5.76  &  4.31  &  5.05  &  4.49  \\
\ti(o)          &  6.10  &  6.07  &  5.97  &  5.55  &  4.83  &  5.71  &  4.89  &  5.22  \\
\ti(i)          &  6.48  &  5.70  &  5.87  &  5.66  &  7.00  &  5.87  &  6.20  &  5.59  \\
\fe(o)          &  ---   &  ---   &   ---  &   ---  &  3.49  &  4.57  &  3.55  &  4.05  \\
\fe(i)          &  4.75  &  3.90  &  4.52  &  3.77  &  ---   &  ---   &   ---  &   ---  \\
\end{tabular}
\end{ruledtabular}
\end{table} 

In order to obtain more detailed insights into the ferroelectricity of
BFTO, the BECs for the different cations, calculated for the fully
relaxed (i.e., without epitaxial constraint) centrosymmetric and
ferroelectric states in both configurations, averaged over similar
(but not necessarily symmetry-equivalent) sites, are presented in
Table~\ref{table:bec}.
It can be seen that the BECs for both \bi\ and \ti\ (and to some
extent also for \fe) are highly anomalous, i.e., they are
significantly increased compared to their formal valences. This
indicates that displacements of the corresponding cations result in
large redistribution of charge and corresponding changes in chemical
bonding,\cite{Ghosez/Michenaud/Gonze:1998,Ederer/Harris/Kovacik:2011}
and is usually considered a signature for ferroelectrically-active
ions.\cite{Posternak/Resta/Baldereschi:1994,Rabe/Ghosez:Book} The most
pronounced difference between the two configurations are the BECs of
the $B$(o) sites in the outer configuration, which are noticeably less
anomalous than those of the $B$(i) sites (and than those of the $B$(o)
sites in the inner configuration). Overall, the BECs for BFTO are
quite similar to the BECs calculated for the related $m=3$ Aurivillius
phase Bi$_4$Ti$_3$O$_{12}$~\cite{Shah:2010kg} and for the related
perovskite system (i.e., $m=\infty$)
BiFeO$_3$.\cite{Neaton_et_al:2005}

The in-plane BECs of \bi\ in the perovskite layer are somewhat more
anomalous than those in the \bio\ layer and also show a more
pronounced reduction when going from the centrosymmetric to the polar
structure. This is consistent with the analysis of Withers {\it et
  al.},\cite{Withers/Thompson/Rae:1991} who suggested the optimization
of bond valence sums of the \bi(perov.) cations as driving force for
the polar displacements in various Aurivillius systems with $m=1,2$,
and $3$. However, a strong reduction of the anomalous in-plane BECs
between the centrosymmetric and polar structures can also be observed
for the \ti(i) sites, which suggests an active role of the inner $B$
site cations for Aurivillius phases with $m>3$.
%
Furthermore, as first pointed out by Perez-Mato et
al.,\cite{Perez-Mato_et_al:2004} the space group symmetry of the
Aurivillius phases allows for coupling between the ferroelectric
distortion and two or more octahedral rotation modes, which can give
rise to so-called \emph{hybrid improper
  ferroelectricity}.\cite{Benedek/Fennie:2011} Indeed, a relation
between the average tolarance factor of the perovskite blocks,
controlling octahedral rotations, and the ferroelectric transition
temperature has been empirically found by investigating a large number
of Aurivillius systems.\cite{Suarez/Reaney/Lee:2001}
It thus appears that ferroelectricity in the Aurivillius phases can
arise from several factors, which cooperate to give rise to the robust
ferroelectric properties observed in this class of compounds.



To summarize, our calculations indicate that it is indeed possible to
control the site preference of the \fe\ cations in BFTO by epitaxial
strain. Tensile epitaxial strain is expected to increase the
occupation of the inner sites with \fe, whereas compressive strain
will lead to a preferential \fe\ occupation of outer sites. In
addition, epitaxial strain also provides an efficient way to enhance
(or reduce) the magnitude of the spontaneous electric polarization,
which, furthermore, is rather insensitive to the actual $B$-site
cation distribution.

We point out that the possibility to tailor site preference in BFTO
also allows to achieve an essentially homogeneous distribution of
\fe\ cations, i.e., no site preference, under moderate compressive
strain. This case could indeed be most favorable for achieving good
percolation of magnetic couplings between the \fe\ ions and thus
promoting long range magnetic order. Therefore, controlling the cation
distribution through epitaxial strain can provide a way of controlling
the multiferroic properties of BFTO and related Aurivillius systems.

\begin{acknowledgments}
This work was funded by ETH Zurich and the Swiss National Science
Foundation under project no. 200021\_141357. We thank Nicol\'o Fanelli
and Rolf Homberger for performing some of the initial calculations for
this work as part of their undergraduate research project at ETH
Z\"urich.
\end{acknowledgments}

\bibliography{references}

\end{document}